\documentclass[twocolumn]{aastex63}

\usepackage{float}
\newcommand{\kepler}{\textit{Kepler}}
\newcommand{\hdone}{HD\,100180}
\newcommand{\hdtwo}{HD\,143761}

\journalinfo{The Astrophysical Journal Letters, accepted}

\shorttitle{A Magnetic Morphology Shift in Old Stars}
\shortauthors{Metcalfe et al.}

\begin{document}

\title{LBT/PEPSI Spectropolarimetry of a Magnetic Morphology Shift in Old 
Solar-type Stars\footnote{The LBT is an international collaboration among 
institutions in the United States, Italy and Germany. LBT Corporation 
partners are: The University of Arizona on behalf of the Arizona Board of 
Regents; Istituto Nazionale di Astrofisica, Italy; LBT 
Beteiligungsgesellschaft, Germany, representing the Max-Planck Society, 
The Leibniz Institute for Astrophysics Potsdam, and Heidelberg University; 
The Ohio State University, and The Research Corporation, on behalf of The 
University of Notre Dame, University of Minnesota and University of 
Virginia.}}

\author[0000-0003-4034-0416]{T.S.~Metcalfe}
\affiliation{Space Science Institute, 4765 Walnut St., Suite B, Boulder, CO 80301, USA}
\affiliation{Max-Planck-Institut f\"ur Sonnensystemforschung, Justus-von-Liebig-Weg 3, 37077, G\"ottingen, Germany}

\author[0000-0003-3061-4591]{O.~Kochukhov}
\affiliation{Department of Physics and Astronomy, Uppsala University, Box 516, SE-75120 Uppsala, Sweden}

\author{I.V.~Ilyin}
\affiliation{Leibniz-Institut f\"ur Astrophysik Potsdam (AIP), An der Sternwarte 16, D-14482 Potsdam, Germany}

\author[0000-0002-6192-6494]{K.G.~Strassmeier}
\affiliation{Leibniz-Institut f\"ur Astrophysik Potsdam (AIP), An der Sternwarte 16, D-14482 Potsdam, Germany}

\author{D.~Godoy-Rivera}
\affiliation{Department of Astronomy, The Ohio State University, 140 West 18th Avenue, Columbus, OH 43210, USA}

\author[0000-0002-7549-7766]{M.H.~Pinsonneault}
\affiliation{Department of Astronomy, The Ohio State University, 140 West 18th Avenue, Columbus, OH 43210, USA}

\begin{abstract}

Solar-type stars are born with relatively rapid rotation and strong 
magnetic fields. Through a process known as magnetic braking, the rotation 
slows over time as stellar winds gradually remove angular momentum from 
the system. The rate of angular momentum loss depends sensitively on the 
magnetic morphology, with the dipole field exerting the largest torque on 
the star. Recent observations suggest that the efficiency of magnetic 
braking may decrease dramatically in stars near the middle of their 
main-sequence lifetimes. One hypothesis to explain this reduction in 
efficiency is a shift in magnetic morphology from predominantly larger to 
smaller spatial scales. We aim to test this hypothesis with 
spectropolarimetric measurements of two stars that sample chromospheric 
activity levels on opposite sides of the proposed magnetic transition. As 
predicted, the more active star (\hdone) exhibits a significant circular 
polarization signature due to a non-axisymmetric large-scale magnetic 
field, while the less active star (\hdtwo) shows no significant signal. We 
identify analogs of the two stars among a sample of well-characterized 
\kepler\ targets, and we predict that the asteroseismic age of \hdtwo\ 
from future TESS observations will substantially exceed the age expected 
from gyrochronology. We conclude that a shift in magnetic morphology 
likely contributes to the loss of magnetic braking in middle-aged stars, 
which appears to coincide with the shutdown of their global dynamos.

\end{abstract}


\vspace*{6pt}
\section{Background}\label{sec1}

The coupled evolution of rotation and magnetic activity in solar-type 
stars has been an active area of research since the pioneering work of 
\cite{Skumanich1972}. The availability of reliable stellar ages has always 
been a limiting factor, with the earliest studies relying entirely on the 
Sun and a few young star clusters. The basic picture that emerged was that 
solar-type stars begin their lives with relatively rapid rotation and 
strong chromospheric activity, but that both properties gradually decay 
with the square-root of the age. The Sun was the oldest star with a 
reliable age beyond 2.5~Gyr \citep{Meibom2015} until the {\it Kepler} 
mission began to yield asteroseismic ages for older field stars 
\citep{Mathur2012, Metcalfe2014, SilvaAguirre2015}. This led to the 
discovery of unexpectedly rapid rotation in this sample \citep{Angus2015}, 
which could be understood if magnetic braking becomes much less efficient 
in solar-type stars beyond the middle of their main-sequence lifetimes 
\citep{vanSaders2016}. A coincident shift in the observed properties and 
prevalence of chromospheric activity cycles \citep{Metcalfe2017} strongly 
suggested a magnetic origin for the lower rate of angular momentum loss.

\cite{Metcalfe2016} proposed that the reduced efficiency of angular 
momentum loss in middle-aged stars could be due to a change in the 
magnetic field morphology. Charged particles in a stellar wind are tied to 
the magnetic field lines until they reach the Alfv\'en radius, which is 
largest for the dipole component of the field and progressively smaller 
for higher-order components \citep{Reville2015}. As a consequence of the 
larger lever-arm, most of the angular momentum loss from magnetized 
stellar winds can be attributed to the dipole component of the field 
\citep{See2019}, so a shift in magnetic morphology from larger to smaller 
spatial scales would reduce the efficiency of magnetic braking. With this 
in mind, \cite{Garraffo2018} suggested a change in magnetic complexity as 
a unifying explanation for the persistent fast rotators in young clusters 
and the anomalously fast rotating old \kepler\ field stars.

There are good reasons why we might have expected a magnetic morphology 
shift in middle-aged stars. According to \cite{vanSaders2016}, spin-down 
stalls at a critical value of the Rossby number, when the rotation period 
becomes comparable to the convective overturn timescale. In this regime, 
convection is no longer influenced by substantial Coriolis forces, and the 
pattern of solar-like differential rotation (i.e.~faster at the equator 
and slower at the poles) either becomes uniform \citep{Featherstone2016}, 
or theoretically might transition to an anti-solar pattern 
\citep{Gastine2014, Rudiger2019}. Observationally, two-thirds of the 
sample of \kepler\ targets with constraints on latitudinal differential 
rotation from asteroseismic mode-splittings are consistent with uniform 
rotation, and none are significantly anti-solar \citep{Benomar2018}. 
\cite{Metcalfe2019} suggested that the resulting loss of shear might 
disrupt the production of large-scale magnetic field by the global dynamo, 
explaining the reduction in angular momentum loss and the gradual 
disappearance of activity cycles in stars beyond the middle of their 
main-sequence lifetimes.

\begin{figure} 
\centerline{\includegraphics[angle=270,width=\columnwidth]{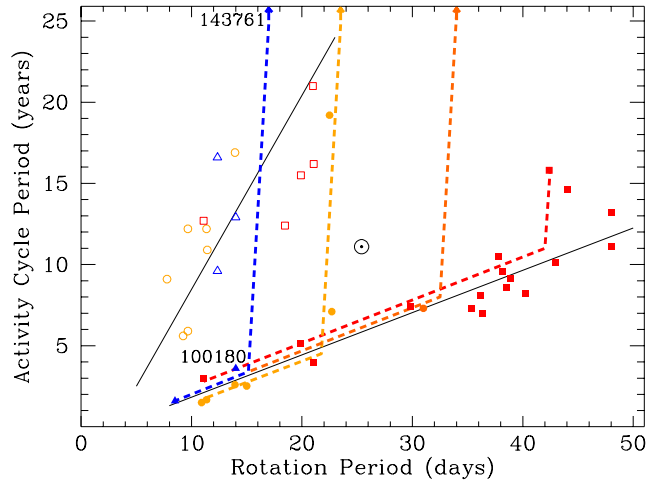}} 
\caption{Dependence of activity cycle period on rotation, showing two 
distinct sequences (solid lines). Points are colored by temperature, 
indicating F-type (blue triangles), early G-type (yellow circles), late 
G-type (orange circles), and K-type stars (red squares). Schematic 
evolutionary tracks are shown as dashed lines \citep{Metcalfe2017}, 
leading to stars with constant activity that appear to have shut down 
their global dynamos (arrows along the top). Our spectropolarimetric 
targets are labeled with their HD numbers.\label{fig1}} 
\end{figure}

We aim to test for the predicted loss of large-scale magnetic field using 
spectropolarimetric measurements of two stars on opposite sides of the 
proposed magnetic transition (see Figure~\ref{fig1}). The more active star 
\hdone\ has a rotation period of 14 days and exhibits dual chromospheric 
activity cycles with periods of 3.6 and 12.9 years 
\citep{Brandenburg2017}. The less active star \hdtwo\ has a rotation 
period of 17 days and shows constant chromospheric activity below the 
solar minimum level over several decades of monitoring at Mount Wilson 
\citep{Baliunas1995, Baliunas1996}. We describe our observations and 
analysis methods in Section~\ref{sec2}. We present the results in 
Section~\ref{sec3}, including an interpretation of the observed Stokes 
profiles and a comparison of the stellar properties with analogs of the 
two stars from the \kepler\ mission. We discuss the results in 
Section~\ref{sec4}, including a prediction of what future asteroseismic 
observations will reveal from the {\it Transiting Exoplanet Survey 
Satellite} \citep[TESS,][]{Ricker2014}.

\section{LBT/PEPSI Observations}\label{sec2}

We observed \hdone\ and \hdtwo\ in May 2019 using the Potsdam Echelle 
Polarimetric and Spectroscopic Instrument 
\citep[PEPSI,][]{Strassmeier2015} at the $2\times 8.4$\,m Large Binocular 
Telescope (LBT) on Mt.\ Graham, Arizona, USA. The two polarimetric units, 
installed at the direct Gregorian focus, were used in circular 
polarization mode with a quarter-wave polymethylmethacrylate (PMMA) 
retarder on a rotary stage in front of the polarizing beam-splitting 
Foster prism unit. The two polarized beams ($I+V$ and $I-V$) are coupled 
with $200\,\mu$m fibers ($1\farcs5$ on sky) to render the light into the 
spectrograph via an image slicer with 5 slices and a resolving power of 
R\,=\,130,000. In polarimetric mode, each spectral order consists of four 
sub-orders with the two polarized beams from each of the two telescopes 
recorded simultaneously. The \'echelle images are recorded on blue 
(480.0-544.1\,nm) and red (627.8-741.9\,nm) channel $10.5\times 10.5$\,k 
STA1600LN CCDs with $9\,\mu$m pixels and 16 amplifiers.

\subsection{Data reduction}

The image processing includes bias subtraction and variance estimation of 
the source images with subsequent super-master flat field correction for 
the CCD spatial noise. Tracing flats are used to define the \'echelle 
orders, scattered light is subtracted from every \'echelle image, and a 
wavelength solution is obtained from the Th-Ar exposures. The optimal 
extraction of image slices and the elimination of cosmic ray spikes is 
then performed for the target image, with subsequent wavelength 
calibration and the merging of slices in each order. Normalization to the 
master flat field spectrum then removes CCD fringes and the blaze 
function. Finally, a global 2D fit is made to the continuum of the 
normalized image, and all spectral orders are rectified into a 1D spectrum 
for a given cross-disperser.

The continuum of the final polarized spectra was further rectified using 
the mean spectrum. The weighted average of all spectra was normalized to 
eliminate any residual effects in the continuum. The ratio of each 
individual spectrum and the mean is then used to fit a smoothing spline, 
which constitutes the improved estimate of the true stellar continuum for 
the individual spectrum.

The polarized spectra were derived with the difference method 
\citep{Ilyin2012} to eliminate any first order residual terms from the 
quarter-wave retarder due to optical misalignment. Two angles on the 
retarder were used to obtain polarization states with opposite sense. The 
difference of the two polarized beams at two angles is combined to obtain 
Stokes $V/I_c$, where $I_c$ is the continuum intensity. The intensity 
spectrum $I$ is the sum of the two polarized beams at two angles. The 
polarized spectra are treated separately for each polarimeter on the two 
telescopes, and are averaged with weights to produce the final spectrum 
\citep[see][]{Strassmeier2019}.

\begin{figure} 
\centerline{\includegraphics[angle=270,width=\columnwidth]{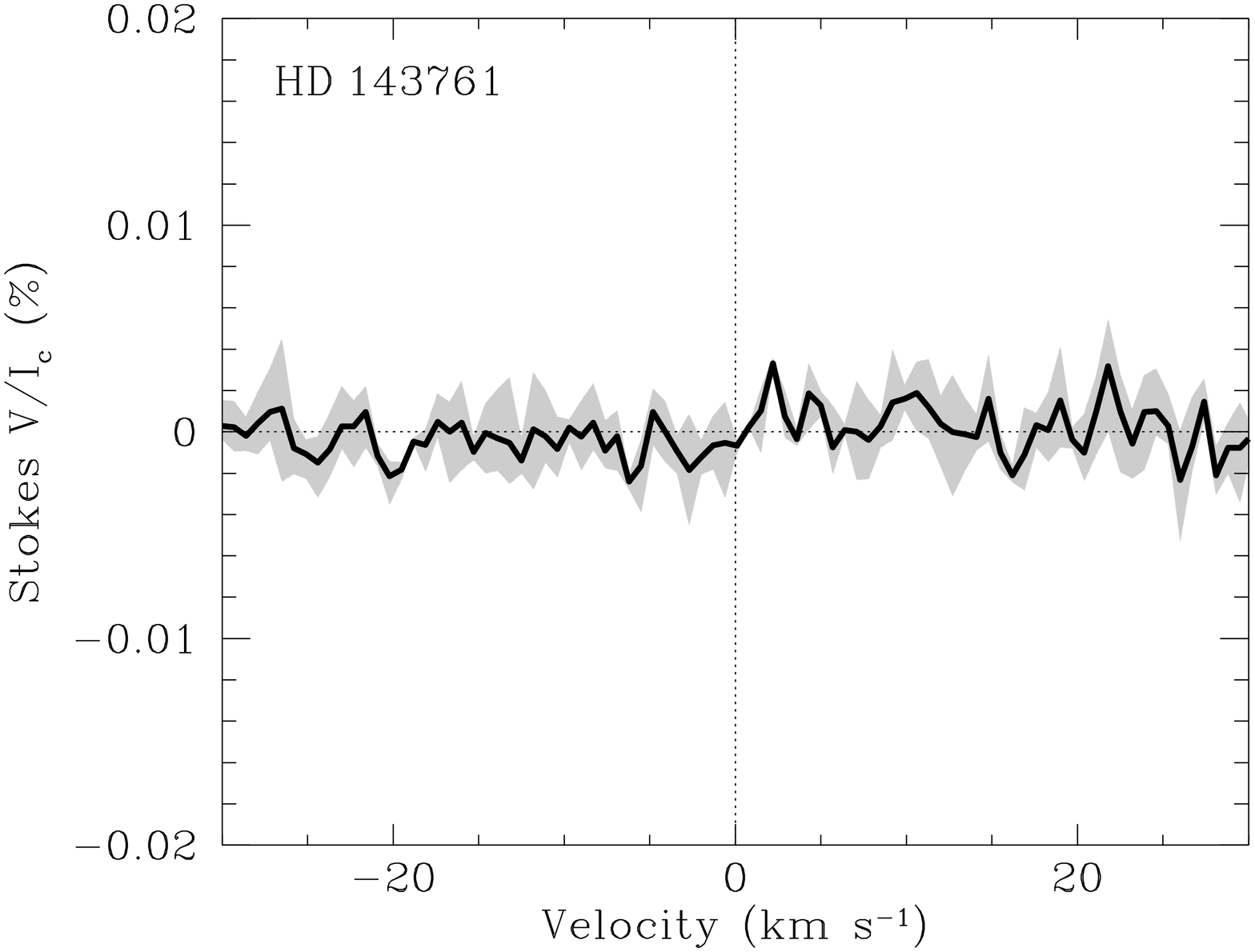}} 
\centerline{\includegraphics[angle=270,width=\columnwidth]{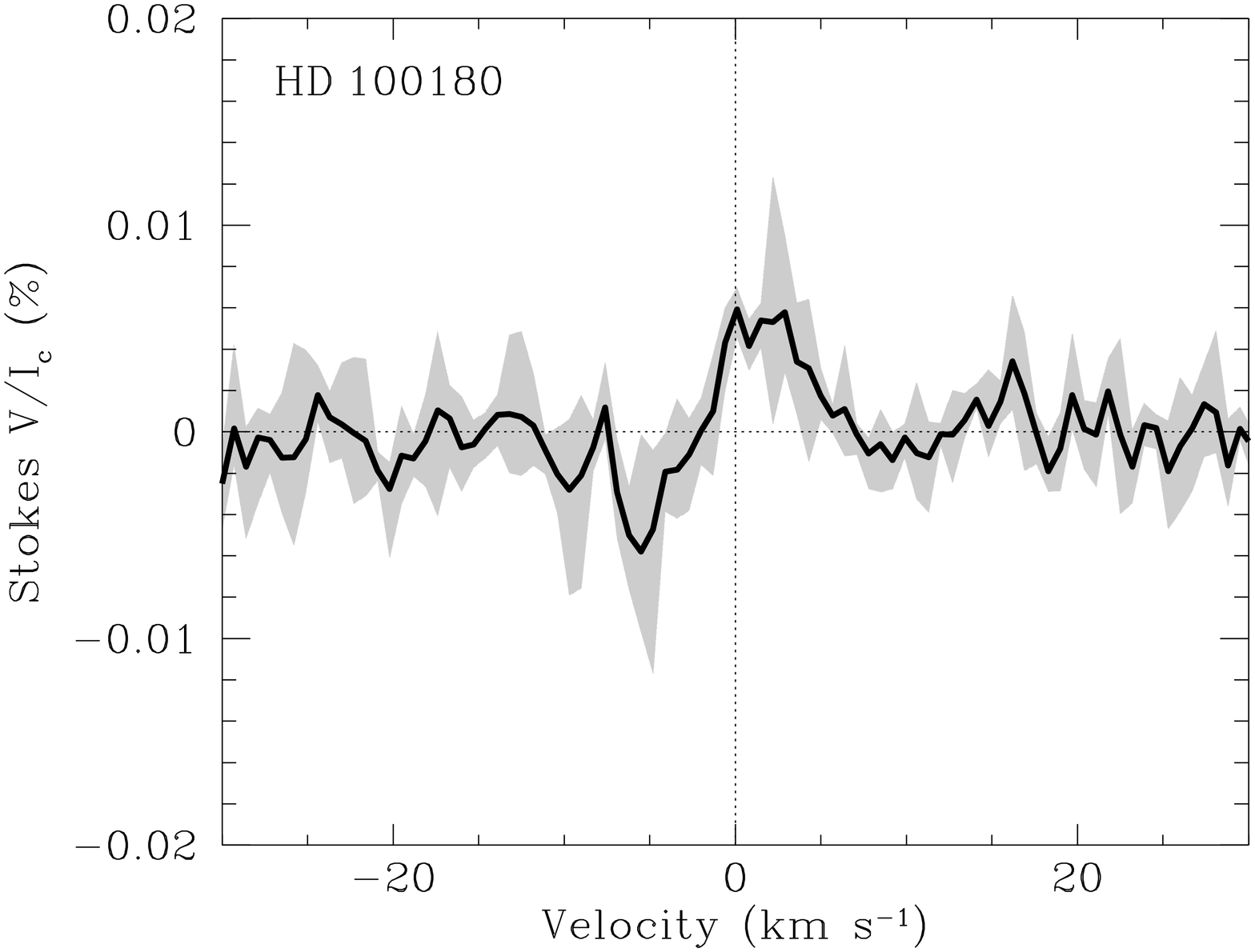}} 
\caption{Least-squares deconvolution of the Stokes $V$ profiles for 
\hdone\ (bottom) and \hdtwo\ (top). The gray shaded regions show the 
range of profiles from individual integrations, while the average profile 
is shown as a dark line. \hdone\ shows the clear signature of a 
large-scale magnetic field, while \hdtwo\ shows no significant 
signal.\label{fig2}} 
\end{figure} 

\subsection{LSD profiles}

Zeeman polarization signatures are typically not detectable in the 
individual spectral lines of even the most active late-type stars. A 
direct detection and quantitative analysis of magnetic field in such cases 
requires the application of some multi-line polarization diagnostic 
method. Here we used the least-squares deconvolution 
\citep[LSD,][]{Donati1997, Kochukhov2010} technique, which derives 
high-quality mean intensity and polarization profiles by weighted 
co-addition of a large number of individual lines. This procedure assumes 
that all line profiles are self-similar and that overlapping lines add up 
linearly. The input line data required by the LSD analysis (central 
wavelengths, effective Land\'e factors and line intensities) were 
retrieved from the VALD database \citep{Ryabchikova2015} using the stellar 
parameters from \citet{ValentiFischer2005}. The final LSD line masks 
employed for \hdone\ and \hdtwo\ included 1040--1180 lines deeper than 
10\% of the continuum. The LSD profiles were calculated with a step of 0.7 
km\,s$^{-1}$, which corresponds to the largest spacing between consecutive 
pixels in our PEPSI spectra.

The LSD procedure was applied separately to the four observations of each 
star (two consecutive integrations with two telescopes) and the resulting 
profiles were inspected for consistency. One out of the four LSD Stokes 
$V$ profiles for \hdtwo\ exhibited enhanced noise due to issues with the 
guiding and wavefront sensors on one telescope, and was excluded from 
further consideration. We have verified that this decision did not change 
any of the conclusions reported below.

The final average circular polarization LSD profiles of both targets are 
shown as dark lines in Figure~\ref{fig2}, with the full range of the 
individual integrations shown as gray shaded regions. The formal 
uncertainty on these mean polarization profiles is 
1.1--1.4\,$\times10^{-5}$, representing a polarimetric sensitivity gain of 
about 50 compared to the original spectra. The LSD Stokes $V$ profile of 
\hdone\ shows a clear polarization signature with a peak-to-peak amplitude 
of about $10^{-4}$. This observation corresponds to a definite magnetic 
field detection, characterized by a false alarm probability (FAP) of less 
than $10^{-10}$ according to chi-square statistics \citep{Donati1992, 
Donati1997}. On the other hand, no evidence of a polarization signal above 
$\approx3\times10^{-5}$ is seen for \hdtwo.

\begin{deluxetable}{lcc}
\tablecaption{Mean Magnetic Field Strengths\label{tab1}}
\tablehead{ \colhead{~~~~~~~~~~~~~~~~~~~~~~~~~~~} & \colhead{~~~HD\,100180~~~} & \colhead{~~~HD\,143761~~~}}
\startdata
$\left<B_{\rm z}\right>$ (G)\dotfill    & $-0.42\pm0.14$ & $-0.27\pm0.12$         \\
$\left<B\right>_{\rm d}$\ (G)\dotfill   & $\cdots$       & $0.48^{+0.76}_{-0.20}$ \\
$\left<B\right>_{\rm q}$\ (G)\dotfill   & $\cdots$       & $1.34^{+2.29}_{-0.45}$ \\
$\left<B\right>_{\rm ZDI}$\ (G)\dotfill & 2.51           & $\cdots$               \\
\enddata
\end{deluxetable}

\section{Interpretation}\label{sec3}

\subsection{Magnetic Field Properties}\label{sec3.1}

We used several methods to characterize the surface magnetic field 
corresponding to the LSD profiles of \hdone\ and \hdtwo. First, we 
measured the mean longitudinal magnetic field $\left<B_{\rm z}\right>$ 
from the first moment of the Stokes $V$ profile \citep{Kochukhov2010}. 
This measurement (see Table~\ref{tab1}) gives the disk-integrated 
line-of-sight magnetic field component. For both stars, $\left<B_{\rm 
z}\right>$ is determined with a precision of $\sim$0.1~G from the PEPSI 
data. For \hdone\ we obtain $\left<B_{\rm z}\right> = -0.42$~G at 
$3\sigma$ significance, confirming the field detection made from the 
Stokes $V$ profiles. For \hdtwo, we obtain a $3\sigma$ upper limit of 
0.36~G.

\begin{figure} 
\centerline{\includegraphics[width=\columnwidth]{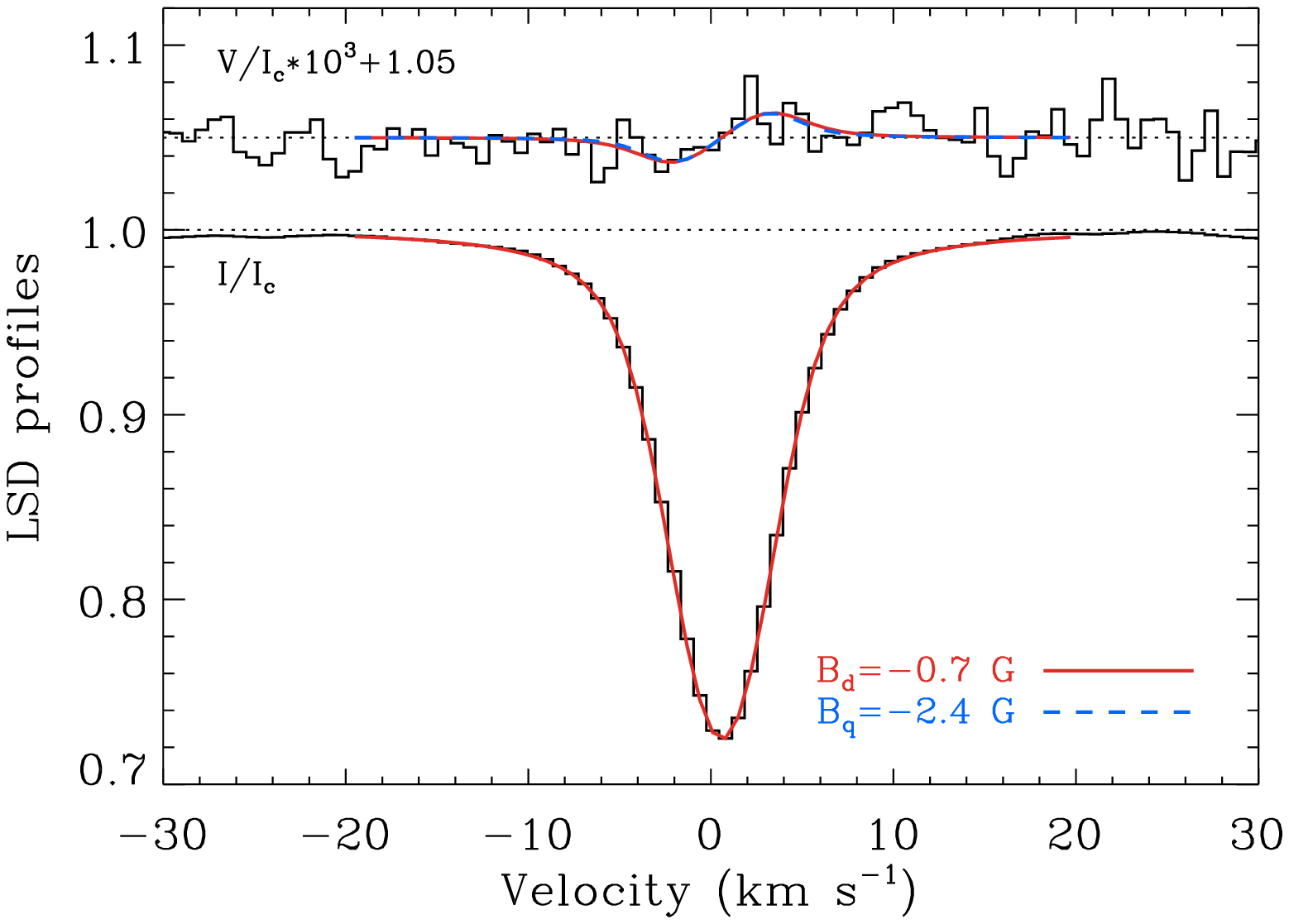}} 
\centerline{\includegraphics[width=\columnwidth]{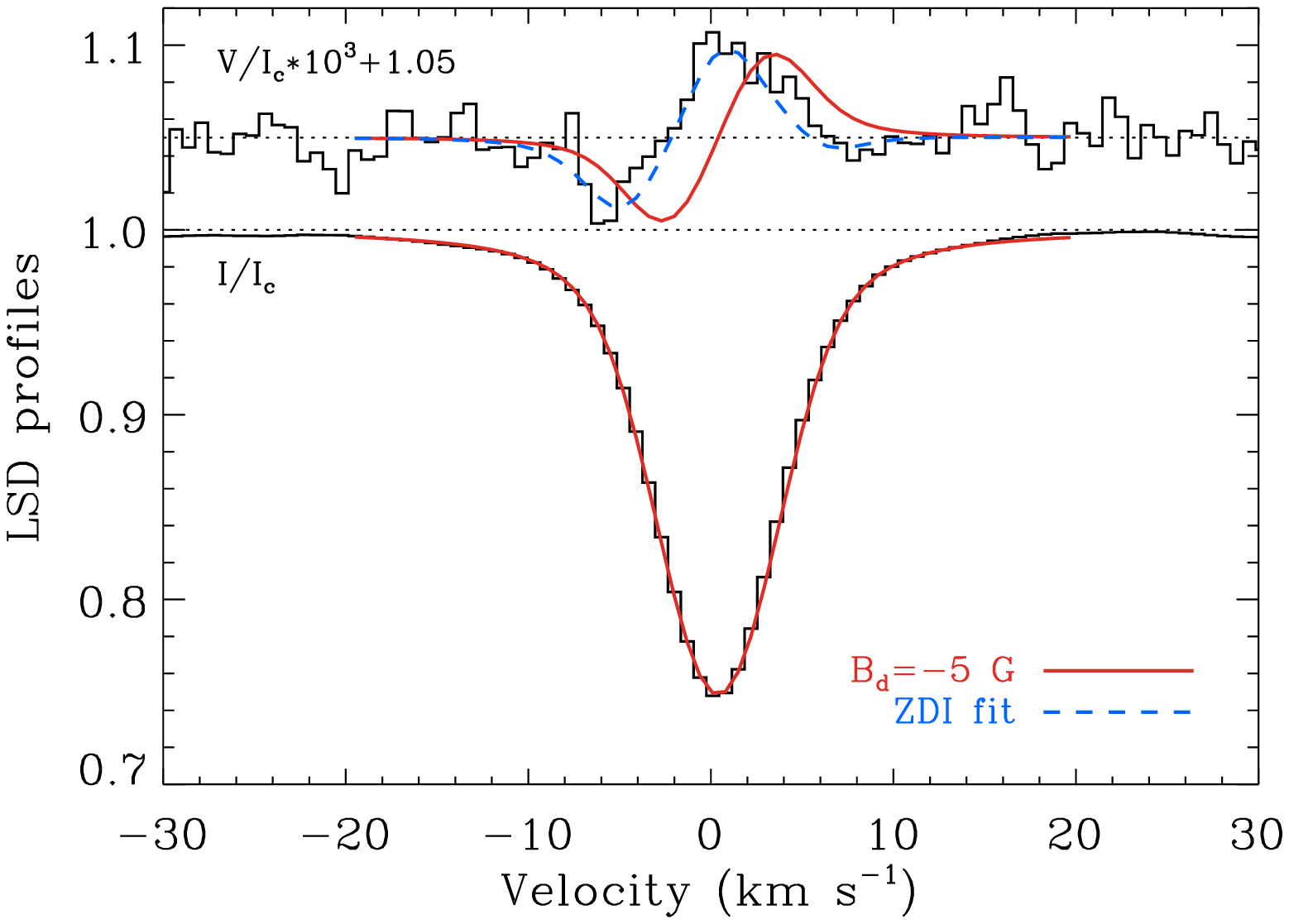}} 
\caption{Models of the observed LSD profiles using fixed inclination and 
various assumptions about the magnetic field morphology for \hdone\ 
(bottom) and \hdtwo\ (top). The offset between observations and the 
dipole model for \hdone\ is due to the non-axisymmetric 
components.\label{fig3}} 
\end{figure} 

Next, we attempted to obtain parameters of the global magnetic field 
morphologies compatible with our observations. Such analysis is best 
carried out using the Zeeman Doppler imaging (ZDI) technique 
\citep{Kochukhov2016}. However, we only have a single-epoch observation 
for each star rather than a full spectropolarimetric time series. 
Therefore, we started with a comparison between the observed LSD Stokes 
$V$ profiles and forward models of the simplest global field 
configurations: axisymmetric dipole and quadrupole fields. We performed 
forward polarized radiative transfer calculations using the ZDI code 
developed by \citet{Kochukhov2014}, employing the analytical 
Unno-Rachkovsky model of the local Stokes parameter profiles. Several 
parameters of this model were adjusted to match the Stokes $I$ LSD profile 
assuming $v \sin i=3.3$ and 1.6 km\,s$^{-1}$ for \hdone\ and \hdtwo\ 
respectively \citep{ValentiFischer2005}. Considering the rotation periods 
reported by \cite{Baliunas1996}, and adopting the stellar radii derived by 
\cite{ValentiFischer2005} for consistency, these $v \sin i$ values suggest 
inclination angles near $i\sim60\degr$ and $i\sim25\degr$ for \hdone\ and 
\hdtwo\ respectively, which we fix for our analysis.

The Stokes $V$ profile of \hdtwo\ is compatible with an axisymmetric 
dipole field with a polar field strength of $B_{\rm d}=-0.7\pm1.1$~G, 
where the error bar corresponds to a FAP\,=\,$10^{-3}$ calculated for the 
difference between the observed and model Stokes $V$ spectra. Similar 
analysis assuming an axisymmetric quadrupole field yields $B_{\rm 
q}=-2.4\pm4.0$~G (see Figure~\ref{fig3}, top panel). These upper limits on 
the polar field strengths can be converted to mean surface field strengths 
of $0.48^{+0.76}_{-0.20}$~G and $1.34^{+2.29}_{-0.45}$~G for the dipole 
and quadrupole morphologies respectively.

\floattable
\begin{deluxetable}{lcc|cc}
\tablecaption{Spectropolarimetric Targets and \kepler\ Analogs\label{tab2}}
\tablehead{\colhead{\hspace*{1.5in}} & \colhead{~~~~HD\,100180~~~~} & \colhead{~~~KIC\,3427720~~~} & \colhead{~~~~HD\,143761~~~~} & \colhead{~~~KIC\,6116048~~~}}
\startdata
B$-$V\ \dotfill                  & 0.57        & 0.55        & 0.60        & 0.59        \\
$T_{\rm eff}$~(K)\ \dotfill      & 5989        & 6043        & 5823        & 6013        \\
$\log g$\ \dotfill               & 4.38        & 4.35        & 4.36        & 4.25        \\
\relax [Fe/H]\ \dotfill          & $-$0.02     & +0.02       & $-$0.14     & $-$0.14     \\
$P_{\rm rot}$~(days)\ \dotfill   & 14          & $14\pm2$    & 17          & $17\pm2$    \\
$\log R'_{\rm HK}$\ \dotfill     & $-$4.92     & $-$4.78     & $-$5.04     & $-$5.02     \\
$t_{\rm gyro}$~(Gyr)\ \dotfill   & $2.1\pm0.4$ & $2.4\pm0.4$ & $2.5\pm0.4$ & $2.7\pm0.5$ \\
$t_{\rm astero}$~(Gyr)\ \dotfill & $\cdots$    & $2.4\pm0.2$ & $\cdots$    & $6.1\pm0.4$ \\
$t_{\rm iso}$~(Gyr)\ \dotfill    & $3.6\pm1.5$ & $3.6\pm0.7$ & $8.4\pm1.7$ & $6.5\pm0.5$ \\
\enddata
\tablerefs{\cite{ValentiFischer2005, Brewer2016, Baliunas1996, Ceillier2016, Garcia2014, Karoff2013, Barnes2007, Creevey2017}.}
\vspace*{-12pt}
\end{deluxetable}

For \hdone\ an axisymmetric dipole with $B_{\rm d}=-5$~G roughly matches 
the observed Stokes $V$ profile amplitude. However, neither dipole nor 
quadrupole axisymmetric morphologies are able to provide an acceptable fit 
to the observed shape of the polarization profile. It appears that the 
surface field geometry of \hdone\ has a dominant non-axisymmetric 
component. In an effort to assess its strength, we let the inversion code 
fit the single Stokes $V$ profile with a general low-order harmonic field 
parameterization usually employed in the ZDI analyses of solar-type stars 
\citep[e.g.][]{Petit2008, Rosen2016}. This calculation yields a 
non-axisymmetric field distribution with a peak local field strength of 
$\approx$\,6~G and a mean field strength of 2.51~G (see Figure~\ref{fig3}, 
bottom panel).

\subsection{Analogs from Kepler}\label{sec3.2}

To place the magnetic properties of \hdone\ and \hdtwo\ in a broader 
context, we searched for analogs of each star within the sample of 
asteroseismic targets observed by the \kepler\ mission. Considering stars 
with detailed asteroseismic modeling from \cite{Creevey2017}, we searched 
for the closest match to both the observed rotation period ($P_{\rm rot}$) 
and the B$-$V color. This procedure identified KIC\,3427720 as the analog 
of \hdone, and KIC\,6116048 as the analog of \hdtwo.

The properties of our target stars and their \kepler\ analogs are listed 
in Table~\ref{tab2}. For our PEPSI targets, the spectroscopic properties 
($T_{\rm eff},\ \log g$, [Fe/H]) come from the analysis of 
\cite{ValentiFischer2005}, while for the \kepler\ analogs we adopt values 
from \cite{Brewer2016}. Rotation periods and chromospheric activity levels 
($\log R'_{\rm HK}$) for our targets were determined by 
\cite{Baliunas1996}. For the \kepler\ analogs, rotation periods were 
determined by \cite{Ceillier2016} and \cite{Garcia2014}, while the 
activity levels were measured by \cite{Karoff2013}. Although KIC\,3427720 
is somewhat more active than \hdone, and KIC\,6116048 is slightly hotter 
than \hdtwo, considering typical uncertainties there is reasonable 
agreement between the stellar properties for each pair.

Using the $P_{\rm rot}$ and B$-$V color for each star, we calculated an 
age and uncertainty following the gyrochronology relation of 
\cite{Barnes2007}. For comparison, we tabulate asteroseismic ages from 
\cite{Creevey2017} for the \kepler\ analogs and isochrone ages from the 
{\tt isochrones} python package \citep{Morton2015} for all of the stars, 
using their spectroscopic properties as input constraints. In order to 
obtain a robust age estimate for each star, we ran {\tt isochrones} 50 
times with each run producing a posterior distribution. We add these into 
a combined posterior distribution for each star, from which we calculate 
the age and its uncertainty from the 50$^{th}$ and 16-84$^{th}$ 
percentiles respectively.

For our more active target \hdone, there is marginal agreement between the 
ages deduced from gyrochronology and isochrones. For the \kepler\ analog 
KIC\,3427720, the ages from gyrochronology and asteroseismology are 
perfectly consistent, while the isochrone age is slightly older. For our 
less active target \hdtwo, there is substantial tension between the ages 
deduced from gyrochronology and isochrones, with the latter suggesting a 
much more evolved state. For the \kepler\ analog KIC\,6116048, both the 
asteroseismic and isochrone ages are considerably older than the age 
suggested by gyrochronology. These results are consistent with the 
suggestion that rotation and activity decouple near middle-age, making 
rotation an unreliable age indicator for older stars \citep{Metcalfe2019}.

\section{Discussion}\label{sec4}

Stars rotate more slowly over time as they lose angular momentum to 
magnetized winds, but most of the resulting torque is exerted by the 
dipole component of the magnetic field \citep{See2019}. Near the middle of 
a star's main-sequence lifetime, the global dynamo that produces 
large-scale field apparently begins to shut down \citep{Metcalfe2017}. The 
result is a decoupling of rotation and magnetism near middle-age 
\citep{Metcalfe2019}, a prediction that can be tested observationally with 
spectropolarimetric measurements. Prior to the transition, more active 
stars like \hdone\ are expected to exhibit clear signatures of a cycling 
large-scale magnetic field, and their rotation periods should be reliable 
age indicators. Beyond the transition, less active stars like \hdtwo\ are 
expected to reach a constant activity state without the largest-scale 
fields that effectively couple rotation and magnetism, so ages from 
gyrochronology will be inconsistent with those derived from other 
techniques.

Our spectropolarimetric measurements of \hdone\ reveal the clear signature 
of a large-scale non-axisymmetric magnetic field with a mean strength of 
2.51~G. By contrast, we do not detect significant polarization in \hdtwo, 
but we can set upper limits on the field strength of a given morphology: 
0.48~G for a dipole field, and 1.34~G for a quadrupole field (see 
Section~\ref{sec3.1}). The chromospheric activity level of \hdtwo\ ($\log 
R'_{HK}=-5.04$), which is sensitive to magnetic heating on all spatial 
scales, suggests that overall it is 76\% as active as \hdone\ ($\log 
R'_{HK}=-4.92$). Considering its sub-solar metallicity, the corrected 
activity level of \hdtwo\ might be slightly lower \citep{Wright2004, 
SaarTesta2012}. This is broadly consistent with the relative strengths of 
$\left<B_{\rm z}\right>$ from our measurements (64\%, see 
Table~\ref{tab1}), although the observed correlation between $\log 
R'_{HK}$ and $\left<B_{\rm z}\right>$ for a large sample of stars shows 
substantial scatter \citep{Marsden2014}. Assuming a fixed morphology, the 
mean dipole field in \hdtwo\ is less than 20\% as strong as the 
non-axisymmetric field in \hdone, while the quadrupole field would be 
closer to the observed ratio (53\%). The available data are consistent 
with the predicted disappearance of dipole field in \hdtwo, but a complete 
ZDI analysis of time-resolved spectropolarimetric observations would be 
required to make this conclusion unambiguous.

The disappearance of large-scale field in stars beyond the middle of their 
main-sequence lifetimes does not apparently create a discontinuity in the 
activity-age relation. Measurements of $\log R'_{HK}$ for a sample of 
spectroscopic solar twins \citep{LorenzoOliveira2018} and \kepler\ 
asteroseismic targets \citep{Metcalfe2016, Booth2019} both show a smooth 
evolution across the range of activity levels where the transition to 
smaller spatial scales is expected to occur ($\log R'_{HK}\approx-4.95$), 
despite a strong discontinuity in the rotation-age relation 
\citep{Metcalfe2017}. Although this may initially seem surprising, it is 
understandable considering the weakness of the global dipole field 
relative to the smaller scale features that mostly cancel in 
spectropolarimetric measurements. For example, when ZDI maps are 
synthesized for the Sun-as-a-star, the dipole component of the field has a 
mean strength $\la$1~G \citep{Vidotto2016}. This can be compared to a mean 
strength of $\left<B\right>\sim170$~G for the unstructured quiet Sun 
\citep{Danilovic2010}, which dominates the contributions to $\log R'_{HK}$ 
because the polarity of the field is irrelevant for chromospheric heating. 
Consequently, disruption of the large-scale organization of the magnetic 
field as the global dynamo begins to shut down can eliminate the dipole 
field with no discernible impact on the activity-age relation.

An additional test of our interpretation will be possible when TESS 
obtains asteroseismic observations of \hdtwo\ in mid-2020. This star is 
currently scheduled to be observed at a 2-minute cadence for up to 54 days 
between April and June 2020, with a high probability of detecting 
solar-like oscillations \citep{Schofield2019}. Given the large difference 
between the ages from gyrochronology and isochrones, and considering the 
asteroseismic age of the \kepler\ analog KIC\,6116048, we predict that the 
asteroseismic age of \hdtwo\ from TESS will be substantially older than 
expected from its rotation period. By contrast, when TESS observes \hdone\ 
in February-March 2020, we expect the asteroseismic age to be consistent 
with gyrochronology.

\acknowledgments The authors would like to thank Matthias Rempel, Keivan 
Stassun and Regner Trampedach for helpful exchanges. This work benefitted 
from discussions within the international team ``The Solar and Stellar 
Wind Connection: Heating processes and angular momentum loss'' at the 
International Space Science Institute (ISSI). T.S.M.\ acknowledges support 
from a Visiting Fellowship at the Max Planck Institute for Solar System 
Research and from the U.S.\ National Science Foundation under grant 
AST-1812634. O.K.\ acknowledges support by the Knut and Alice Wallenberg 
Foundation (project grant ``The New Milky Way''), the Swedish Research 
Council (projects 621-2014-5720, 2019-03548), and the Swedish National 
Space Board (projects 185/14, 137/17). K.G.S.\ thanks the LBTO for the 
flexible PEPSI scheduling, and the German Federal Ministry BMBF and the 
Brandenburg State Ministry MWFK for funding PEPSI. D.G.R.\ and M.H.P.\ 
acknowledge support from NASA ADP grant 80NSSC19K0597.


\end{document}